\documentclass[aps,prb,twocolumn,superscriptaddress,floatfix,showpacs]{revtex4}
\usepackage{graphicx}
\usepackage{amssymb}
\usepackage[usenames]{color}

\begin{document}

\title{A cluster model with random anisotropy for hysteresis jumps in CeNi$_{1-x}$Cu$_{x}$ alloys}

\author{J. R. Iglesias}\affiliation{Instituto de F\'{\i}sica, Universidade Federal do Rio Grande do Sul, 91501-970 Porto Alegre, Brazil}
\author{J. I. Espeso}\affiliation{Dept. CITIMAC, University of Cantabria, 39005 Santander, Spain}
\author{N. Marcano}
\affiliation{Dept. CITIMAC, University of Cantabria, 39005 Santander, Spain}\affiliation{ICMA, CSIC-Universidad de Zaragoza, 50009 Zaragoza, Spain}
\author{J. C. G\'omez Sal}\affiliation{Dept. CITIMAC, University of Cantabria, 39005 Santander, Spain}

\begin{abstract}
Some Cerium compounds exhibit hysteresis cycles with sharp macroscopic jumps in the magnetization at very low temperatures. This effect is attributed to the formation of clusters in which the anisotropy competes with the applied magnetic field. Here, we present a simple model where a lattice of ferromagnetically coupled spins is separated in clusters of random sizes and with random anisotropy. Within this model, we obtain hysteresis cycles presenting jumps that behave in a similar way that the experimental ones, and that disappear when increasing the temperature. The results are in good agreement with the hysteresis cycles measured at very low temperatures in CeNi$_{1-x}$Cu$_{x}$ and the comparison with these experimental results allows to discriminate the relative importance of the mechanisms driving the thermal evolution of the cycles.
\end{abstract}

\pacs{71.27.+a, 75.60.Ej, 61.46.+w, 75.10.-b}

\maketitle

\section{Introduction}
The increasing importance of ``inhomogeneous'' magnetic systems is due to the fact that inhomogeneities (phase separation, local impurities, etc.) are on the basis of new emergent magnetic phenomena such as colossal magnetoresistance (CMR)~\cite{Burgy,deTeresa} or non conventional superconductivity.~\cite{Mathur} However, these systems also provide, as it is clearly the case of magnetic clustering,~\cite{Binder} a direct link between the magnetic properties of mesoscopic entities and the bulk macroscopical observations.

Substitutional compounds, in which disorder effects are intrinsically present, have been used to tune some attractive magnetic effects such as Kondo lattice,~\cite{Coqblin} Non Fermi Liquid State~\cite{Stewart} or quantum criticalities.~\cite{Gegenwart} In particular, some well extended theories~\cite{Miranda} consider systems consisting of ``magnetic clusters'' embedded in a ``non magnetic'' or ``paramagnetic'' matrix (Griffiths phases), to describe the Non Fermi Liquid State~\cite{CastroNeto}. One of the more representative examples of such situation is the compound CeNi$_{1-x}$Cu$_{x}$.~\cite{Marcano05} For this system it has been proposed~\cite{Marcano07} a spin cluster-glass state which, by means of a progressive percolative process obtained by lowering the temperature, finally reaches a long-range ferromagnetic state at low temperatures.

This behavior has been confirmed by two kinds of experiments, Small Angle Neutron Scattering (SANS) measurements, which demonstrates the existence of ``magnetic clusters'' of around 20\AA ,below the freezing temperature, and low temperature hysteresis cycles, obtained in the ferromagnetic state, presenting steps on the magnetization. These steps have been attributed to the avalanches of domain flips, as a mesoscopic analogue of the Barkhausen noise,~\cite{Lhotel,Tung} in priority to other possible mechanisms proposed in the literature.~\cite{Mahendiran,Friedman,Caneschi} In this case, it was proved that the ``percolated clusters'', reaching a minimum energy situation, displays the ``magnetic domain'' structure existing in a conventional ferromagnet.

\begin{figure}[t]
\centerline{\includegraphics[width=6.5cm, clip=true]{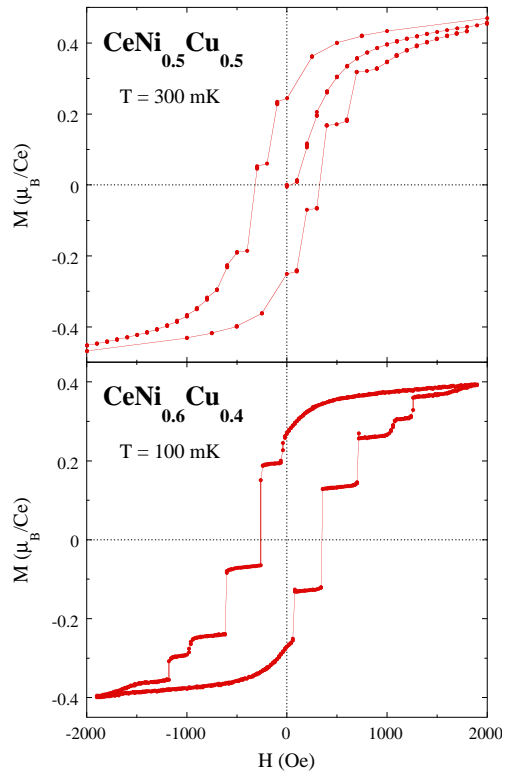}}
\caption{Hysteresis cycles of two representative compositions (CeNi$_{0.5}$Cu$_{0.5}$  and CeNi$_{0.6}$Cu$_{0.4}$) measured at very low temperatures. Note that the jumps take place upon reversal of the magnetic field and the number of jumps is the same in both branches of the hysteresis loop.}
 \label{fig1}
\end{figure}

The main characteristics of the observed hysteresis loops have recently been presented in an experimental article~\cite{Marcano} and can be summarized as follows. The steps are extremely sharp at the lowest measured temperatures ($\sim$ 100 mK) and strongly dependent on temperature; i.e., at 300 mK the steps disappear in the CeNi$_{0.6}$Cu$_{0.4}$ compound. The steps appear after the sign inversion of the magnetic field and the number of jumps is the same in both branches of the hysteresis cycle (increasing or decreasing the magnetic field). In Fig.~\ref{fig1} we present the hysteresis cycles for two characteristic ferromagnetic compounds (CeNi$_{0.5}$Cu$_{0.5}$ and CeNi$_{0.6}$Cu$_{0.4}$). The ferromagnetic long-range order state has been confirmed by neutron diffraction at these low temperatures.~\cite{Espeso,GomezSal}
At this point, we also have to remind that the compositional evolution along the series tunes different physical interactions, such as the Kondo one, which gets enhanced when approaching CeNi and also, in particular, the RKKY interactions, which evolve from the antiferromagnetic behavior of CeCu to a ferromagnetic one when decreasing the Cu content.~\cite{Espeso} This experimental result supports the competition of positive and negative magnetic interactions for the intermediate compositions and, thus, this competition has to be accounted for  in the model subsequently developed.

The hypothesis used to explain the emergence of jumps in the low temperature hysteresis cycles in this system is that magnetic domains developed from a percolative process of static ferromagnetic clusters, reaching a minimum energy state at low enough temperatures.~\cite{Marcano07} This mechanism increases the magnetic correlation length up to values that can be detected by neutron diffraction ($\sim$ 10$^3$ \AA). The process is driven by the increasing importance of the RKKY interaction as the temperature decreases. This interaction, then, competes with the local anisotropy, and gives rise to a structure of magnetic domains that displays an ``asperomagnetic'' mesoscopic state such as the one reported by Coey~\cite{Coey} in the case of amorphous systems. The present situation is clearly reminiscent of that case, but occurring in crystalline samples. The initial conditions for such behavior are: random anisotropy for the clusters, disorder and competing magnetic interactions between and inside the clusters. A large number of magnetic compounds present special features dealing with inhomogeneous states, associated to spin or cluster-glasses, phase separations, etc.~\cite{Burgy,Ohno,Uehara} The incidence of these features on their magnetic behavior is of great importance in determining and characterizing new ground states of matter (Quantum Critical Points, etc.) at low temperatures.

A first attempt to theoretically describe the magnetization jumps in the hysteresis cycles of the CeNi$_{1-x}$Cu$_{x}$ alloys was previously developed by using a spin Hamiltonian, including a spin-spin interaction term, an applied external field and a random local field.~\cite{Garcia} This random local field was intended to simulate the effect of a local anisotropy, but instead of considering just an axis of easy magnetization, there was also a preferred sense, what means that the anisotropy energy was minimum when the spin was aligned with the local random field, and maximum when pointing in the opposite sense. The hysteresis cycles so obtained presented a number of jumps strongly dependent on the value of the spin, (assumed equal or bigger than one half), and on the number of possible discrete orientations for the random field. Even if this model describes the qualitative features of the phenomena, the jumps in the hysteresis cycle are equally spaced, as both the spin and the anisotropy can only have discrete values. Moreover, in that case, the anisotropy was locally described at each site. Thus, the model is not well suited to represent a system of clusters.

We propose here an alternative model that seems to be better adapted to the description of the studied effects. On one hand, this model is simpler than the previous one, being done that now the spins are described by an Ising model and have just two possible orientations. On the other hand, anisotropy can now have random orientation, and clusters of random different sizes with different anisotropy directions are considered. Then different clusters have different anisotropy orientation, but inside each cluster all spins are subjected to the same anisotropy direction, which is not related to the cluster size. This way of describing the anisotropy field directly arises from the model proposed to describe the magnetic behavior of the CeNi$_{1-x}$Cu$_{x}$ compounds.\cite{Marcano07} and is a useful tool to describe bulk properties arising from a microscopic dynamics. Finally, the temperature effects are not restricted to the thermal activation of the spins, but a phenomenological description of the percolative process is also considered. We also remark as a difference with the previous model that each cluster is now characterized by a true anisotropic term; i.e., by an axis of easy magnetization, whose direction is random, and not by a random field, as in ref. ~\cite{Garcia}.
We should recall at this point that, although the random anisotropy model developed in the present work provides a description closer to the empirical point of view, there exists much further literature on random field models, including extensive reviews.\cite{Belanger,Sethna} However, it is also true that both kind of models are closely related\cite{Fedorenko} and, even, both of them belong to the same universality class.\cite{Dahmen}

Therefore, we have undertaken a numerical study to model the hysteresis cycles of CeNi$_{1-x}$Cu$_{x}$ as a function of temperature and we present here the results of our Montecarlo calculations. In the next section we will describe the model and the simulations, while in section III we present and discuss the results.

\section{Description of the Model}

The model we consider is a three dimensional Ising system on a simple cubic lattice. The lattice is separated in clusters of random mesoscopic size, and each cluster is characterized by a random anisotropy direction. The interaction between the spins is mainly ferromagnetic but we allow a given concentration of antiferromagnetic interactions to represent the disorder and the fact that the magnetic order is different in the limit of high Ni or high Cu concentrations. Then, the Hamiltonian of the system reads:
\begin{equation}
H = -\frac{1}{2}\sum_{i,j}J_{ij} S_iS_j-\sum_n \sum_{i \Subset n} (S_i
 A \cos{\phi_n})^2-\mathcal{H}\sum_i S_i
\end{equation}
\noindent where $S_i$ are Ising spins that can take the values $\pm1$, so describing a system with spin $1/2$. This situation is representative for a Ce$^{3+}$ ion in an orthorhombic environment, as its J = 5/2 ground state is split into three doublets and, thus, only one doublet is relevant at very low temperatures. The spins interact among them through the magnetic interactions $J_{ij}= \pm J$, the plus sign representing ferromagnetic interactions and the minus sign antiferromagnetic ones. The value of the exchange constant is used in the present work as the scaling factor to present the results on other magnitudes; i.e., the anisotropy, the temperature and the magnetic field will be measured in units of $J$. We also consider that the lattice is divided in clusters of random sizes, and that each cluster, $n$, is characterized by an anisotropy field $\mathbf{A_n}$ whose absolute value is assumed to be constant, $A$, while the orientation with respect to the $z$-axis, $\phi_n$, is random. We remark at this point that the anisotropy strength and direction is assigned to each individual spin, and that the calculations are made on the basis of these individual spins. Therefore, the energy barrier for spin flip is related only to the local strength of the anisotropy, $A \cos{\phi_n}$, the exchange energy, $J$ (times the number of neighbors), and the applied field, $\mathcal{H}$. So, differently to the previous models (including ref. \onlinecite{Garcia}) here the system is separated in clusters of random sizes, and the second term correspond to an anisotropy interaction characterized only by the direction (and not the sense) of the anisotropy axis. Moreover, we are trying to describe a disordered compound in which there is a strong competition between positive and negative magnetic interactions. We must recall here that CeCu is antiferromagnetic~\cite{Espeso} while a ferromagnetic ordered phase builds up at low temperatures for the Ni-richer alloys, as a consequence of a percolating process of magnetic clustering.~\cite{Marcano05, Marcano07} Hence, as we are interested in the hysteresis cycles at very low temperatures, we assume that in the Hamiltonian (1) the interactions $J_{ij}$ are mainly ferromagnetic, $J_{ij}= +J$, and we introduce a small, but finite, concentration of AF links, $J_{ij}= -J$.

\begin{figure}[t]
\centerline{\includegraphics[width=8cm, clip=true]{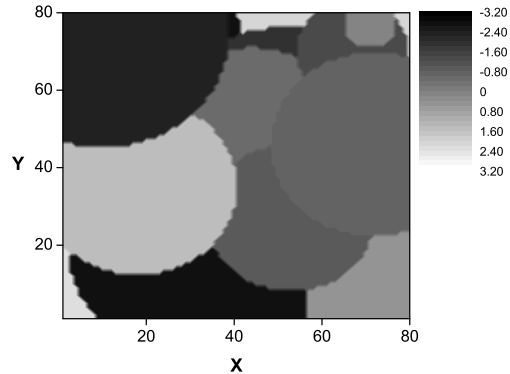}}
\caption{Two dimensional section of the lattice showing the clusters and the relative angle of the anisotropy, between $-\pi$ and $-\pi$. For this figure we use $r_{min}=0.2N$ and $r_{max}=0.5N$}
\label{fig2}
\end{figure}

Then, within this model, we have performed a Monte Carlo simulation on a 3-dimensional lattice, at zero and finite temperatures, considering a cubic lattice of dimension $ N \times N \times N$ with $N=80$. We considered a number of clusters of the order of $N$, because we have observed that a good qualitative description of the experimental results is obtained for relatively big clusters and a small dispersion in size. The strength of the anisotropy has been taken as $A=3J$, and the random orientation between $0$ and $\pi$. Then, clusters are generated according to the following recipe: a) an arbitrary number of spherical clusters $R$ is assumed, being $R$ of the order of $N$, b) the center of each cluster is determined at random within the sample, c) The radius of the cluster is chosen at random with uniform probability in a range $\{r_{min},r_{max}\}$, d) to prevent superposition of clusters, sites are labeled as belonging to a cluster once its anisotropy is determined. Eventually some sites may be out of any cluster, but the number of such sites not belonging to any cluster is much smaller than the total number of sites. To illustrate the cluster geometry, in Fig.~\ref{fig2} we show a section of a lattice exhibiting a typical $2D$ cluster distribution. It is evident that some big clusters are present and they are the ones that determine the strength of the jumps in the hysteresis cycle. In the present simulation we have taken $R=N$ and relatively big clusters, $0.25 \leq r_{min} \leq 0.45$ and $0.4 \leq r_{max} \leq 0.65$.

\begin{figure}[t]
\centerline{\includegraphics[width=8cm, clip=true]{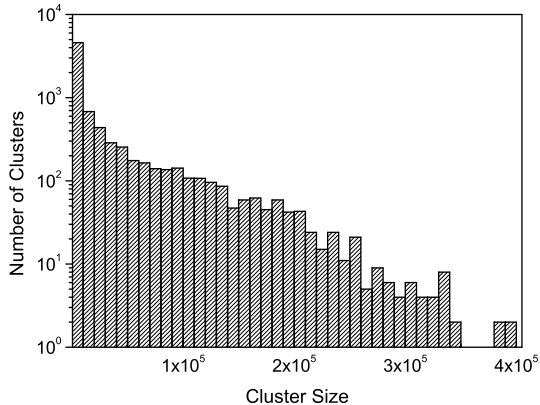}}
\caption{Distribution of the size of the clusters (measured as the number of sites of the cluter) evaluated over $400$ samples, with the same parameters of Fig.~\ref{fig2}}
\label{fig3}
\end{figure}

 We have also determined the cluster size distribution. As the samples are rather small to perform a statistics, we have studied  $400$ samples, and the results are plotted on Fig.~\ref{fig3}. The distribution is represented in a semi-log plot, and the behavior is almost an exponential law. The average size of the clusters is $18$ sites but with a relatively high variance, of the order of $44$. That means that there are a big number of small clusters filling the space among a few large clusters. Those large clusters are enough to cover most of the lattice, and we can see from the figure that the frequency of large clusters (size bigger than $10^5$ sites) is of the order of the number of samples.

We should keep in mind that, as the anisotropy term in the Hamiltonian (1) is quadratic, the contribution of this term to the total energy is independent of the spin being $S=\pm1$. This fact implies that, if one considers a dynamics leading to the state of minimum energy, there would be no effect of this term. Hence, we assume that, in order to change the spin direction in one site, the system has to overcome an energetic ``barrier''; i.e., it needs to have enough energy to outdo the strength of the anisotropy field. Therefore, at zero temperature, we consider that if the spin is in a state of negative energy, it will remain there independently of any consideration about a possible (more negative) energy of the state with opposite spin. Just when the energy becomes positive, the spin will reverse its direction.
Finally, as it was stated above, a percentage of the links $J_{ij}$ are taken to be antiferromagnetic. In the results presented in this work, this percentage is considered to be $5\%$.

Two different processes that influence the temperature evolution of the jumps observed in the hysteresis cycles will be considered:

a) Thermal activation: This is the usual process where the spin always changes its direction if its energy is positive but, even if the energy of the spin $S_i$ on the $i$-site is negative, $-E$, we assume that there is a finite probability $\frac{1}{2}e^{-E/T}$ that the spin changes its orientation.

b) Cluster percolation: In this case, we will have to account for this effect in a qualitative way, as we do not have a precise mathematical model on the temperature evolution of the percolation paths. However, what we know from the experimental results,\cite{Marcano07} is that the percolation process yields an increase of the ferromagnetic correlation length when lowering the temperature. Thus, in our calculation, we can associate the different stages of this percolating process with different cluster size distributions; each of the distributions  centered on larger sizes as the temperature goes down.

\section{Results and Conclusions}

\begin{figure}[t]
\centerline{\includegraphics[width=6cm, clip=true]{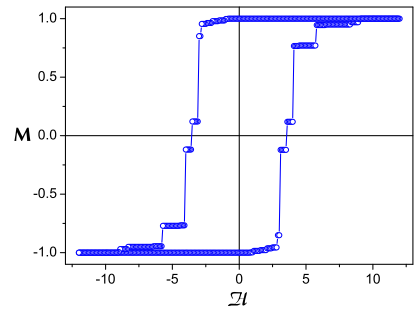}}
\caption{One example of the hysteresis cycles obtained by numerical simulation, with the parameters described in the text. The magnetic field is measured in units of $J$ and the magnetization is relative to the saturation value.}
\label{fig4}
\end{figure}
\begin{figure}[t]
\centerline{\includegraphics[width=8cm, clip=true]{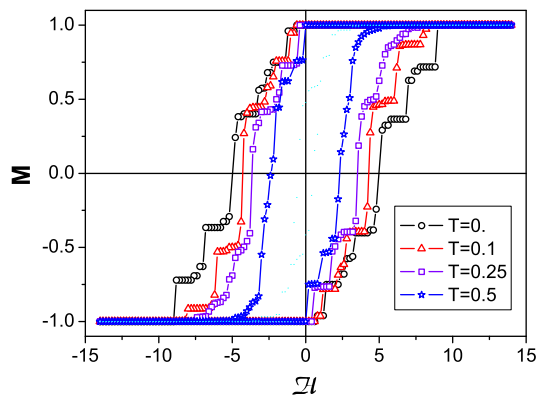}}
\caption{Effect of temperature activation only: Variation of a simulated hysteresis cycle as a function of temperature. The steps practically disappear for a temperature of the order of $T \sim 0.5J$. The parameters (size of the lattice, anisotropy, disorder, size distribution of the clusters) are the same as those used for obtaining Fig.~\ref{fig4}, but different seeds have been taken for the random number generator.}
\label{fig5}
\end{figure}
In Fig.~\ref{fig4} we present a typical hysteresis cycle obtained at $T=0$ using the model described above. The hysteresis cycles obtained with the simulation exhibit some remarkable coincidences with the experimental results (see Fig.~\ref{fig1}). Not only the number of jumps is the same when inverting the magnetic field but the cycle presents the same inversion symmetry with respect to the origin, characteristic of the experimental results. And those features are independent of the number of jumps obtained by using different seeds for the random number generator in the simulation. Also, the disorder introduced with the elimination of some ferromagnetic links produce a curvature of the magnetization curve for low values of the magnetic field, $\mathbf{\mathcal{H}}$, that qualitative coincides with the experimental results and was not observed in previous calculations.~\cite{Garcia} This qualitative agreement between simulation and experiments strongly suggest that the  existence of clusters with different anisotropy directions is at the origin of the jumps in the hysteresis cycles, as well as other properties of the CeNi$_{1-x}$Cu$_{x}$ compounds. Each of the jumps of the hysteresis cycle corresponds to an avalanche process where the spins (of one or more clusters) align with the applied magnetic field. The bigger the size of the cluster, the bigger the avalanche and the lower the number of jumps. So, one can conclude from the simulations that the clusters are of significant, mesoscopic size, as if one reduces too much the size of them, the number of jumps should be much larger than in the experimental results. We remark that the number of jumps is so low that it is not possible to make an statistical analysis, and if one makes statistics over different samples the size of the jumps is so different among different samples that the effect would disappear. However, we have performed simulations with different size of the samples and/or the clusters, and with different seeds for the random number generator and the results are consistent with the above mentioned conclusions

\begin{figure}[t]
\centerline{\includegraphics[width=7cm, clip=true]{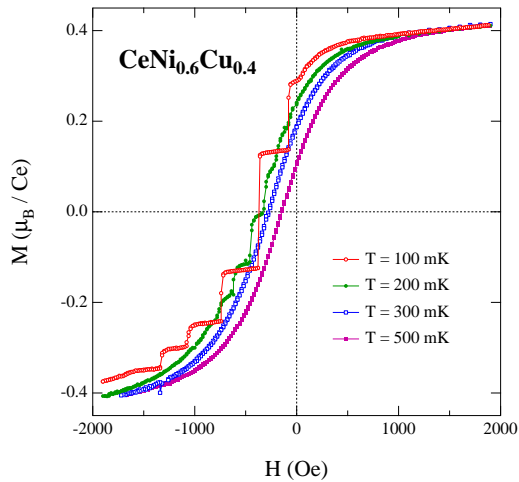}}
\caption{Temperature evolution of the left branch of the hysteresis cycle for a representative composition (CeNi$_{0.6}$Cu$_{0.4}$). Note that the coercive field almost does not change whilst the jumps are still noticeable, and it clearly decreases for higher temperatures, once the jumps have disappeared.}
\label{fig6}
\end{figure}

We have also performed finite temperature calculations. First we have just consider thermal activation: an example of the temperature variation of the hysteresis cycles is presented in Fig.~\ref{fig5}, where one can see that the jumps in the hysteresis cycle disappear as the temperature increases, in very good agreement with the experimental data, presented in Fig.~\ref{fig6}. In the simulation, the temperature is measured in units of $J$. So, comparing with the experimental results, where the jumps are fully wiped out at temperatures of the order of $500mK$,  we can estimate the value of the exchange interaction as being of the order of $1K$. This is a too small value for the exchange interaction, but this result can be attributed to the fact that in the simulation the number of spins in each cluster is several orders of magnitude smaller than in the physical system. In addition, we should keep in mind that this system exhibits some characteristics of Kondo behavior, so that the value of the magnetic moments at very low temperatures are considerably reduced, as observed by neutron diffraction.~\cite{Espeso,GomezSal}

\begin{figure}[t]
\centerline{\includegraphics[width=8cm, clip=true]{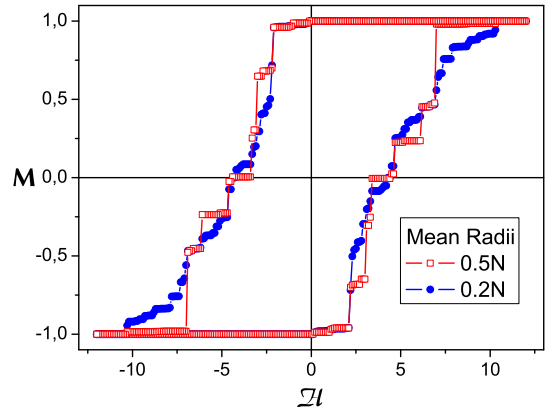}}
\caption{Effect of the cluster percolation on the hysteresis cycles. The different stages of the percolation process are correlated with cluster distributions around different sizes: the larger the size, the lower the temperature. The rest of the parameters (size of the lattice, anisotropy, disorder) are the same as those used for obtaining Fig.~\ref{fig5}.}
\label{fig7}
\end{figure}

Then, we have tried to show the effect of the percolation process. As stated before, the different stages of this percolation process can be qualitatively correlated with cluster distributions centered around different sizes: the lower the temperature, the larger the size of the clusters. These qualitative results are presented in Fig.~\ref{fig7} and the observed behavior shows clearly that the jumps are smoothed as the mean radii of the clusters is reduced; i.e., the temperature is increased. This result is an indication that the cluster percolation process is also able to account for the temperature evolution of the experimental hysteresis cycles.

Although ab-initio considerations arising from the proposed model do not allow to clearly distinguish which one of the mechanisms is more relevant in order to define the variation with the temperature, the analysis of the results concerning both of these mechanisms may give us relevant information in this sense. If we look at the coercive field variation arising from both situations, we will realize that it is much larger when considering the thermal activation process than when taking into account the percolation one. Comparing now with the experimental results presented in Fig.~\ref{fig6}, we can observe that the coercive field do not hardly change whilst the jumps in the hysteresis cycle are still noticeable (T $<$  300 mK), whereas it strongly decreases with the increasing temperature once the jumps have completely disappeared. This fact is a clear indication that the percolative process is the dominant one at the lowest temperatures, but once a ``conventional'' hysteresis cycle is reached, the thermal activation becomes predominant.

So, keeping in mind the limitations of the present simulation, performed with a relative small number of spins, the model proposed  here can well account for the striking properties of the hysteresis cycles of the CeNi$_{1-x}$Cu$_{x}$ compounds, describing both the jumps of the magnetization at zero temperature and the extinction of the jumps when the temperature increases. A number of facts indicates that the model presented reflects in an appropriate way the experimental evidence and the proposed scenario of percolative ferromagnetic clusters for this series:
\begin{enumerate} \vspace{-3mm}
\item [a)] The estimated low value for the exchange energy is a signature of the Kondo interaction. \vspace{-3mm}
\item [b)] Some amount of disorder in the magnetic interactions is necessary to account for the curvature of the magnetization curves in the low field range. \vspace{-3mm}
\item [c)] The clusters must be of mesoscopic size in order to obtain a small number of jumps, as observed experimentally. \vspace{-3mm}
\item [d)] The formation of these clusters of mesoscopic size is intrinsically related to the disorder in the interactions and the small value of the magnetic moments.
\item [e)]The temperature evolution of the hysteresis cycles is dominated by the percolation process at the lowest temperatures and by the thermal activation at higher ones.
\end{enumerate} \vspace{-3mm}
These points are critical for an understanding of the ferromagnetic cluster percolative model. In similar series based in other Rare-Earths, such as Nd or Tb, no evidence of cluster creation has been found.~\cite{Senas} Therefore, we can conclude that the calculations presented here provide a remarkable support for the model proposed in references \onlinecite{Marcano05} and \onlinecite{Marcano07}, which, in fact, could be a valid description for a large number of strongly correlated electron systems with disorder effects.

This work was supported by Brazilian agencies CNPq and FAPERGS through Pronex 04/0874.9, the Spanish MAT 2003-06815 project and the European COST P16 program.

\bibliography{Clusters}

\end{document}